\documentclass{article}

\usepackage{arxiv}

\usepackage[utf8]{inputenc} 
\usepackage[T1]{fontenc}   
\usepackage{hyperref}    
\usepackage{url}           
\usepackage{booktabs}   
\usepackage{amsfonts}      
\usepackage{nicefrac}     
\usepackage{microtype}     
\usepackage{lipsum}		
\usepackage{graphicx}
\usepackage[numbers]{natbib}
\usepackage{doi}
\usepackage{amsmath}
\usepackage{alltt}
\usepackage{listings}
\usepackage{tcolorbox}
\usepackage{makecell}
\usepackage{float}\usepackage{array}
\title{Generative AI in Heritage Practice: Improving the Accessibility of Heritage Guidance}

\date{31 August 2025}	

\author{ 
    \href{https://orcid.org/0000-0003-1144-8065}{\includegraphics[scale=0.06]{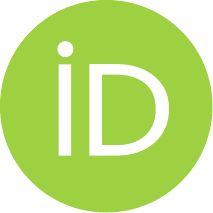}\hspace{1mm}Jessica Witte} \\
    Edinburgh Futures Institute\\
    University of Edinburgh\\
    \And
    \href{https://orcid.org/0000-0002-1166-0878}{\includegraphics[scale=0.06]{orcid.pdf}\hspace{1mm}Edmund Lee} \\
    Historic England\\
    \And
    Lisa Brausem\\
    Historic England\\
    \And
    Verity Shillabeer \\
    Historic England\\
    \And
    \href{https://orcid.org/0000-0002-0872-0614}{\includegraphics[scale=0.06]{orcid.pdf}\hspace{1mm}Chiara Bonacchi} \\
    School of History, Classics \& Archaeology\\
    University of Edinburgh\\
}

\renewcommand{\shorttitle}{\textit{arXiv} Template}

\hypersetup{
pdftitle={A template for the arxiv style},
pdfsubject={q-bio.NC, q-bio.QM},
pdfauthor={David S.~Hippocampus, Elias D.~Striatum},
pdfkeywords={First keyword, Second keyword, More},
}

\begin{document}
\maketitle

\begin{abstract}
This paper discusses the potential for integrating Generative Artificial Intelligence (GenAI) into professional heritage practice with the aim of enhancing the accessibility of public-facing guidance documents. We developed HAZEL, a GenAI chatbot fine-tuned to assist with revising written guidance relating to heritage conservation and interpretation. Using quantitative assessments, we compare HAZEL's performance to that of ChatGPT (GPT-4) in a series of tasks related to the guidance writing process. The results of this comparison indicate a slightly better performance of HAZEL over ChatGPT, suggesting that the GenAI chatbot is more effective once the underlying large language model (LLM) has been fine-tuned. However, we also note significant limitations, particularly in areas requiring cultural sensitivity and more advanced technical expertise. These findings suggest that, while GenAI cannot replace human heritage professionals in technical authoring tasks, its potential to automate and expedite certain aspects of guidance writing could offer valuable benefits to heritage organisations, especially in resource-constrained contexts.
\end{abstract}

\section{Introduction}
\label{sec:introduction}

Generative artificial intelligence (GenAI) is increasingly integrated into cultural heritage practice, but the study of such uses is still in its infancy. This article investigates the extent to which fine-tuned GenAI models offer new possibilities for enhancing the accessibility and readability of heritage-related guidance, working specifically with documents published by Historic England (HE). Since the GenAI chatbot ChatGPT was released by the company OpenAI in 2022, ‘generative artificial intelligence’ has often been referred to as ‘AI’ or ‘artificial intelligence’ in scientific literature \cite{leey24}, \cite{dahmen23}  and colloquially \cite{ciriello25}, \cite{forlini23}, \cite{ng25}. Though ChatGPT and similar technologies are a type of AI, we refer to specifically generative AI systems and technologies as ‘GenAI’ and the broad field of artificial intelligence as ‘AI’. 

Artificial intelligence (AI) has been deployed in cultural heritage practices related to preservation, dissemination, and conservation. For example, researchers have used AI for automating the classification of soil colours according to the Munsell system \cite{milotta20}  and for recognising handwriting in medieval manuscripts \cite{cilia20}.  In galleries, libraries, archives, and museums (GLAM), mass digitisation projects facilitated by the development of optical character recognition (OCR) technologies have benefitted from AI-augmented improvements in the scanning technologies themselves, along with more robust post-OCR corrective pipelines \cite{hegg22}. Additionally, AI-based handwritten text recognition (HTR) software such as Transkribus has been widely adopted for digitisation in archives and libraries \cite{nockels22}. The increasing availability of digitised materials has, in turn, made possible large-scale digital cultural heritage projects which have
explored library and archive metadata, social media data, and other forms of cultural data \cite{underwood20}, \cite{jones24}. This ‘big data’ heritage research, which employs text mining, natural language processing \cite{bonacchi24},  and other computational quantitative approaches, highlights a significant methodological shift in heritage that departs from deeply more traditionally qualitative analysis \cite{bonacchi19}.   
 
The development of GenAI has been similarly reshaping heritage research and practice. Though examples of real-world implementations are still fairly limited due to the novelty of these technologies, GenAI has been used in information retrieval and dissemination tasks \cite{cobb23}.  For instance, galleries and museums have integrated GenAI into the structure of their exhibits to tailor the interpretation of collections for various audiences, such as visitors with disabilities \cite{shehade23}.  In these contexts, GenAI systems guide visitors through exhibits, tell stories, and even provide recommendations \cite{trhicho23}. In libraries, AI-powered chatbots have been developed to assist with obtaining relevant information \cite{lappa23} and increasing the accessibility of collections \cite{verma23}.  GenAI has also been used for restoring and reconstructing heritage objects, such Roman coins \cite{altaweel23}  and works of art \cite{obrien23}.  
 
However, despite increasing numbers of use cases for GenAI in heritage settings, researchers have yet to fully outline the potential impacts of this emerging technology on the sector. This is concerning, since high-profile data privacy \cite{wagner21}, \cite{hollingshead21}, and AI scandals \cite{wolf17}, \cite{dastin22}, have already demonstrated the importance of thoroughly attending to any potential risk factors before implementing new technologies. While our study does not claim to speak to the full spectrum of possible risks, we do aim to outline the major affordances and constraints of GenAI when integrated into the process of developing heritage guidance documents. To the best of our knowledge, our study is the first to explore GenAI specifically for this purpose, and is significant because heritage guidance informs the curation of objects, ideas and places from the past and, in turn, the construction of memory. 

In the next sections, we discuss GenAI and the role it can play as a writing assistant. Thereafter we introduce HAZEL, a chatbot powered by a fine-tuned GPT 3.5-turbo model that was developed to assist HE authors in drafting and revising guidance documents. We then present the combination of quantitative methods used to assess HAZEL’s outputs in a series of heritage guidance writing and revision tasks. Results derived from four established readability formulas reveal modest improvements over the GPT 3.5 model behind ChatGPT at the time of development. However, manual assessments performed by a copyeditor highlight limitations in HAZEL’s cultural sensitivity and ability to address the technical complexities often inherent in heritage documentation. We conclude with recommendations for responsibly introducing GenAI in heritage contexts through, for instance, alignment with FAIR (Findable, Accessible, Interoperable, and Reusable) data principles. 

\section{GenAI ‘intelligence’ and capabilities as writing assistant 
}
\label{sec:background}
\subsection{Artificial intelligence and cognition}
For decades, AI researchers have debated the potential for machine or computational intelligence. Much of this discourse has focused on the distinction between ‘narrow AI’—systems designed to perform specific tasks \cite{kurzweil05}—and ‘artificial general intelligence’ (AGI), which aspires to match or exceed human capabilities in a wide range of tasks across domains \cite{adams12}, \cite{mitchell24} Though hypothetical, many prominent GenAI developers position AGI as central to their mission. OpenAI underscores the importance of ensuring that ‘artificial general intelligence benefits all of humanity’ in its research and development \cite{openabout25}. On his personal blog, OpenAI CEO Sam Altman claims not only that 'we are now confident we know how to build AGI as we have traditionally understood it', but also that AGI is a precursor to 'superintelligence' \cite{altman25}.

Yet this optimistic vision of AGI—and of AI’s broader trajectory—is far from universally accepted. Many experts argue for the ontological impossibility of AGI, contending that machines, at best, simulate rather than embody cognition. They lack key features of human intelligence, such as contextual and analytical awareness, emotional understanding, and subjective experience \cite{sokolowski88}, \cite{dreyfus67}, \cite{dreyfus92}, \cite{dreyfus13}, \cite{landgrebe22}, \cite{searle80}, \cite{shuering24}, \cite{fjelland20}. For example, the Large Language Models (LLMs) through which GenAI systems function have been criticised for lacking any lived experience or coherent mental model of the data on which they are trained. In other words, no matter how sophisticated, GenAI systems ultimately amount to a ‘data dump of ones and zeroes’ \cite{vallor24}. As a result, critics argue, these technologies demonstrate a lack of self-awareness, as evidenced by their frequent inability to recognise or correct errors in their generated outputs \cite{kaomi24}, \cite{pushpanathan23}.  

In contrast, other studies suggest that LLMs indeed demonstrate cognitive capabilities based on their performance on well-known benchmarks. For instance, recent research shows that LLMs can pass the Turing Test \cite{jones25} and score highly on standard intelligence assessments such as the Wechsler Adult Intelligence Scale \cite{galatzer24}.  However, intelligence tests may not be sufficient for estimating cognition in computers because these evaluations primarily measure a narrow subset of abilities associated with pattern recognition, memory, and rule-based reasoning. These are tasks that lend themselves to convergent thinking, or the ability to arrive at a single correct answer using existing knowledge, rather than divergent thinking, which involves generating multiple novel or unconventional solutions to open-ended problems \cite{guilford67}.  LLMs excel at convergent thinking tasks because they largely operate through probabilistic pattern-matching—what some researchers call ‘stochastic parroting’\cite{bender21}  or ‘statistical autocompletion’ \cite{fayyad23}.  Humans, by contrast, use both convergent and divergent thinking in creative problem solving and other tasks \cite{brophy01}.  

\subsection{Potential risks of GenAI as Writing Assistant for the heritage sector}
GenAI technologies have already been informally integrated into professional practice across various sectors, including heritage \cite{alnaqbi24}, \cite{spennemann24}. However, we found that the application of GenAI for writing and editing heritage-focused texts—such as policy, guidance, and interpretation—has been under-investigated. This may be due to the novelty of the technologies themselves, as well as the lower emphasis on computational tools in some areas of heritage practice. LLMs reflect the biases embedded in their training data in which certain languages, actors and perspectives are dominant; this can lead to the systems amplifying certain power imbalances embedded in cultural and historical records \cite{ray23}. If integrated at scale into the writing process for heritage outputs, GenAI systems could therefore reinforce existing dominant-culture biases in policy and guidance. Such risks are compounded not only by the very rapid pace of development in the so-called ‘AI boom’ \cite{tang20}, \cite{strickland21} but also by what some former OpenAI staff see as developers’ failure to prioritise safety and safeguarding \cite{samuel24}. 

While the heritage sector is only beginning to examine these implications, a growing body of research has already identified a range of risks associated with GenAI. These include concerns about intellectual property rights \cite{lucchi24}, \cite{quang21}, embedded biases \cite{hacker24}, \cite{kaplan24}, \cite{motoki24}, \cite{warr24}, privacy breaches \cite{etminani24}, and environmental impact \cite{dong24}, \cite{ding25}. Troublingly, biases have also been observed in the GenAI system’s decision making when it makes assumptions about users based on their demographic information such as their race \cite{kanhai24}.  For example, one recent study found that ChatGPT tailored its recommendations of programmes of study based on undergraduate students’ demographic profiles \cite{zheng24}.  

Ethical concerns are not limited to GenAI chatbots’ outputs, but can arise throughout the entire research and deployment pipeline \cite{wu23}. Although some developers of LLMs have introduced safety protocols and guardrails, users can ‘jailbreak’ or circumvent them through structured prompts \cite{zeng24}, \cite{shen24} that coax systems into generating misleading, inaccurate, unethical, or even illegal content \cite{nah23}. Even when used in good faith, GenAI remains prone to generating misinformation. In scholarly writing tasks, for example, LLMs frequently fabricate bibliographic citations that appear credible but do not exist \cite{spennemann23}, \cite{buchanan24}. 

Another growing area of inquiry concerns the impacts of GenAI on writers and the writing process. In various studies, GenAI systems have performed competitively on convergent thinking tasks, or those with one correct answer, such as multiple-choice exams, proofreading, and structured problem-solving \cite{sallam24}, \cite{orru23}, \cite{castgonz22}. However, their performance declines significantly on divergent tasks, which require originality, creativity, or critical insight. For instance, essays ‘written’ by ChatGPT lack evidence of advanced critical thinking, nuanced argumentation, and depth of exploration into topics \cite{pavlik23}. Researchers have also found that, when asked to write a series of stories, ChatGPT produces formulaic narrative arcs in its own predictable writing style \cite{doshi23}, \cite{zhuoyan24}. Thus, a hypothetical use case of GenAI for fiction writing at scale could potentially erode the diversity of authorial voices and narratives in the cultural record, thereby contributing to the homogenisation of literary works. 

In addition to concerns about the potential effects of GenAI on texts themselves, researchers have also considered how working with these novel technologies might impact authors. Research on this topic has examined GenAI’s impact on self-efficacy, or writers’ independence in overcoming challenges that occur throughout the writing process \cite{washington23}. This work has concluded that although GenAI can augment self-efficacy by assisting with brainstorming \cite{li24},  story creation \cite{pellas23},  and basic editing \cite{koltovskaia24},  habitually relying on GenAI assistants can impede writers’ development of problem-solving skills \cite{washington23}. This is especially a concern when the writing task at hand extends beyond the capabilities of GenAI, such as for divergent tasks requiring expert-level knowledge or creativity \cite{gao25}. 

Taken together, current research highlights that while GenAI tools offer functional benefits in writing tasks, their limitations pose significant risk. In the heritage sector, where written outputs such as policy documents and interpretive guidance shape both professional practice and public understanding, these risks are compounded by longstanding concerns about representational equity and accessibility. To better understand how GenAI could impact heritage communication in practice, it is useful to examine how these texts are currently written and disseminated.

\subsection{Writing Heritage Guidance: the case of Historic England}
In the UK, organisations that oversee heritage sites publish materials outlining best practices for conservation along with guidelines informing planning decisions. In England, Historic England (HE), which ‘helps people care for, enjoy and celebrate’ the country’s ‘historic environment’ \cite{herole25},  has published over 270 of these documents in its ‘Advice and Guidance’ collection \cite{heguidance25}.  These materials explain how readers can ‘care for and protect historic places’ as well as ‘understand how heritage fits into the wider planning agenda’ \cite{headvice25}. Guidance production at HE has redirected focus on a codification of sector expert knowledge towards a role in sharing expertise with a wider audience including the public \cite{lee13}. This is recently exemplified in the revision of web pages, as a part of HE's climate action strategy. For example, the 'Your Home' web pages present advice for owners of older or listed buildings \cite{hehome25}. Heritage guidance materials published by HE therefore have a variety of potential audiences including heritage professionals, homeowners, city planners, community members, government officials, local councils, architects, and anyone involved with decision-making related to conservation. 

Depending on topic and scope, guidance documents can be lengthy, complex, and technical even for their intended target audiences. Another factor affecting ease of communication is that most of the guidance has been published in PDF format. Although PDFs allow for agile sharing and are secure, as they cannot be immediately altered, they are not generally machine-readable. Therefore, readers who use assistive technologies such as text-to-speech software may be unable to access the text in PDF files \cite{pradhan22}. Accessibility issues in PDFs are especially heightened in cases in which tables, diagrams, or images are present, which is the case for much of the technical guidance published by heritage organisations such as Historic England \cite{sorge20}.  

While Historic England’s guidance plays a central role in shaping conservation practice across England, its effectiveness depends not only on the accuracy of its content but also on whether it can be understood by a wide and varied readership. The following section draws on scholarship from writing studies and education to explore how written communication functions across different audiences.

\section{Written communication and heritage practice}
\label{sec:context}
\subsection{Writing for different audiences: accessibility and readability}
Researchers and educators have been grappling with the challenge of determining the extent to which written content can be understood by various audiences. In an increasingly interconnected world, experts in fields such as writing studies and education have embraced writing assessment practices that emphasise fairness over accuracy and exactness \cite{cushman16}, \cite{elliot16}. This approach, which considers factors such as whether a writer is using their native language, whether their background includes well-resourced academic institutions, and whether they have a disability, is, by nature, a subjective evaluation. 

Additional attention has been directed to the subjective nature of obstacles that readers might experience. These include sensory, cognitive, motor, language, expertise, cultural, and media impediments and, sometimes, a combination \cite{maas20}, \cite{rink24}. Addressing communication barriers can help to ensure that writing is accessible; put another way, accessible writing aims to understand ‘the barriers that prevent access to the content’ and how they can be ‘removed’ \cite{maas20}. 

One widely recommended strategy for eliminating these barriers is the use of plain language, which means avoiding slang, jargon, idioms, or other complex terms. Plain language can help to ensure that information communicated is more accessible to target audiences, especially if these are not sector specialists \cite{boldyreff01}, \cite{maas20}. However, the revision process can be complex and challenging, possibly involving multiple steps requiring both the original author and an editor to review a text \cite{kirkpatrick17}. This process could therefore be difficult to implement in organisations with limited resources or skills.

In contrast to subjective approaches focused on fairness and accessibility, other researchers have developed objective frameworks to evaluate the readability of texts. For instance, experts have created pipelines that aim to quantify objectively–often computationally– the reading difficulty of a particular text through characteristics such as word count, sentence length, and the number of concrete versus abstract words \cite{zedelius19}, \cite{pires17}. These computational frameworks originate from a series of formulas for defining and assessing the readability of written work, a term which refers to the ease of understanding a particular text for general audiences \cite{lorge44}, \cite{klare74}, \cite{fry68}, \cite{dubay04}, \cite{bormuth69}. Using features such as the average sentence length, the average number of syllables per word, and the difficulty of words used, these formulas score reading difficulty on scales that correspond to the predicted level of educational attainment required to easily understand a text \cite{flesch48}, \cite{dale49}, \cite{smith67}. Therefore, more concise texts typically score as more readable than texts with longer sentences and words.

While critiques of readability scores point out the limitations of quantifying something which, to most readers, is intuitive \cite{irwin80}, subjective \cite{coke73}, and, when calculated using formulas, only ever \textit{predictive} \cite{klare74}, the Flesch Readability ease and Flesch-Kincaid formulas have been widely applied to assess the readability of technical writing, healthcare communication, and web content design. In comparable studies, these metrics have been implemented to assess readability of GenAI-produced content \cite{nikolayeva24}, \cite{rouhi24} and descriptive annotations, or comments written in files of code \cite{eleyan20}. Based on these formulas, GenAI has been demonstrated to improve readability of technical writing disseminating public health information \cite{abreu24}, \cite{roster24}. However, other studies following a similar research design found that ChatGPT-produced technical writing scored as ‘difficult’ to ‘very difficult’, or readable for audiences with university-level education \cite{momenaei23}, \cite{gencer24}.  

Although accessibility and readability both estimate how effective a particular text will be at addressing a general audience, we refer to accessibility as a qualitative summary of the subjective, person-centred barriers a reader could face in understanding a text. Accessibility, therefore, considers not only the text itself but also the media through which it is made available (and to whom), in addition to other factors such as compatibility with assistive technologies, colour scheme, layout, and so forth. Readability, on the other hand, is an estimate of reading difficulty based on a quantitative assessment of its text-level features. In other words, readability provides an insight into the accessibility of a text’s content that is limited to estimating the challenge for a general audience to understand what it communicates. Readability scores, which measure the complexity of sentences and words, can provide objective assessments of a text’s accessibility–at least in terms of whether its diction, syntax, or structure present potential barriers to its readers.

These distinctions between accessibility and readability are particularly important when applied to written outputs in the heritage sector, where guidance materials and interpretive texts must address diverse and often non-specialist audiences. 

\subsection{Readability \& Accessibility in Heritage Writing}
Though comprehensive automated solutions have yet to be proposed for revising heritage guidance for readability and accessibility, the capabilities of GenAI technologies offer a potential pathway worth assessing. In heritage specifically, a growing body of literature has examined how meaning is communicated in museum contexts, particularly through interpretation, exhibit labels, and digital content \cite{ravelli96}, \cite{ravelli06}, \cite{lazzeretti16}. Although not concerned with heritage guidance specifically, this literature highlights the existence of several possible issues that may hinder the accessibility of heritage texts. For instance, written interpretation often contains jargon, complex syntax, and passive language, even when exhibiting characteristics of text deemed ‘readable’ by existing metrics, such as short sentences \cite{kjeldsen15}. Other characteristics of interpretation labels themselves, such as writing style, legibility, and complexity, can also affect their readability and, in turn, visitors’ interest in reading them \cite{screven92}.

A central concern in this work is the accessibility of museum texts for audiences who rely on multimodal interpretation, such as audio descriptions and guides. Yet these forms of interpretation have also been shown to include features that pose communication barriers, such as technical language and complex sentence structures \cite{perego19}, \cite{randaccio22}. Readability formulas that rely primarily on surface-level text features—such as word or sentence length—may under-estimate the actual difficulty of a given passage, particularly if it contains dense jargon or culturally specific references. Therefore, reducing communication barriers in museum texts requires attention to more than what can be quantified linguistically.

In other words, because exhibitions function as ‘interpretive communities’ \cite{hooper20} where a wide range of actors—artefacts, displays, technologies, curators, visitors, and histories—interact \cite{ravelli96}, museum interpretation should be understood as part of a broader assemblage \cite{deleuze87}  rather than as isolated text. More broadly, heritage writing that fails to account for multiple perspectives and entry points risks reinforcing an authorised heritage discourse (AHD) that ‘works to naturalise a range of assumptions about the nature and meaning of heritage’ \cite{smith06}. 

To explore these concerns empirically, the following section outlines our methodological approach to fine-tuning GenAI for improving the accessibility and readability of heritage texts—specifically focusing on guidance produced by Historic England.

\section{Materials \& Methods}
\label{sec:methods}
This study aimed to explore how GenAI technologies could be adapted to support the production of accessible and readable heritage guidance. Our approach focused on assessing, fine-tuning, and evaluating a large language model (LLM) to assist Historic England in improving the clarity and inclusivity of its guidance publications. To do this, we developed HAZEL, a GenAI chatbot powered by an LLM that we fine-tuned to revise heritage guidance content for readability and accessibility standards in addition to Historic England’s editorial requirements.

This section outlines the methods used to design, build, and evaluate HAZEL. We begin by identifying the accessibility and readability criteria that revised guidance should meet, developed in consultation with Historic England’s Content team. We then describe a series of experiments designed to assess the baseline capabilities of ChatGPT, followed by the process of preparing training data and fine-tuning the model. Finally, we explain the testing completed to evaluate HAZEL’s performance, which included both automated readability analysis and qualitative evaluation by a professional copyeditor familiar with Historic England’s house style.

\subsection{Measures of accessibility to capture}
When designing our framework for fine-tuning and assessing the HE guidance corpus and evaluating the performance of HAZEL, we first outlined the global characteristics and values pertaining to accessibility that we would hope to see reflected in HE guidance. To do so, we adapted existing concepts and measures of accessibility discussed in the previous section along with HE’s accessibility guidelines. Through in-depth conversations with the HE Content team, whose main duties involve supporting the drafting, revision, and publication process for guidance and other web content, we agreed that revised guidance should be:
 
\begin{itemize}
    \item Readable, demonstrating a Flesch readability score of at least 50/100;
    \item Concise, with sentences of 20 words or fewer;
    \item Clear, written unambiguously in plain English with minimal use of technical terms;
    \item Professional in tone, e.g. avoiding contractions;
    \item Formatted and written for readers with various abilities; and
    \item Consistent with Historic England’s house style guidelines and general brand values, such as inclusivity, diversity, and equality \cite{heinclusion25}.
\end{itemize}

Using these criteria, we proceeded to first assess the existing capability of ChatGPT, in order to establish whether these would be sufficient, before proceeding to fine-tune ChatGPT to create HAZEL. The quantitative elements of readability and conciseness were measured computationally, while the qualitative elements were assessed by the project team and a copyeditor with expertise in HE content and style.

\subsection{Preliminary assessment of ChatGPT}
LLMs and GenAI systems have been described as black boxes \cite{rohan23}, \cite{franzoni23} because many of these models were not developed transparently and characteristics such as their parameters and training data are unknown. Therefore, we aimed to outline the default behaviours of ChatGPT through a prompting experiment. At the time this research was conducted, ChatGPT was built on a fine-tuned GPT-3.5 model. Transcripts of interactions with ChatGPT can be found in the appendix to this paper. We posed questions to the chatbot intended to highlight both its capabilities and limitations in tasks related to the HE guidance writing process, which we summarise in the Results and Discussion sections of this paper. The findings of this assessment of ChatGPT informed our decision to proceed to fine-tuning the model, but also offer some more general insights into how GenAI systems function. 
 
We first aimed to establish whether we could reliably improve ChatGPT’s performance in writing tasks through prompt engineering, an iterative process of adjusting inputs to optimise chatbots’ outputs \cite{giray23}, \cite{debnath25}. This stage of research involved instructing ChatGPT to complete tasks related to the research and writing process for heritage guidance. Here, we provided the criteria for revised HE guidance that we had agreed upon in consultation with the Content team as overall instructions for the chatbot to follow, and then proceeded to prompt ChatGPT to revise excerpts from published guidance according to these guidelines. We also asked ChatGPT a series of related questions designed to test its baseline performance in a variety of relevant tasks including its knowledge of British spelling and grammatical conventions, understanding of (and ability to produce) readable content, and writing and editing abilities. 
 
Additionally, we tested ChatGPT’s ability to estimate readability by asking it to calculate the Flesch-Kincaid and Flesch Readability Ease scores of two excerpts taken from guidance documents published by Historic England. We compared ChatGPT’s assigned scores to those of popular readability software Readable.com and Microsoft Word’s built-in readability statistics. Flesch-Kincaid and Flesch Readability Ease formulas quantitatively estimate the difficulty of texts based on factors such as the number of syllables per word and the number of words per sentence. The first excerpt was a passage of 245 words from the ‘3D Laser Scanning for Heritage’ guidance document, which consists of 119 pages, 15 of which include references, a glossary, and a list of resources \cite{helaser18}.  The second excerpt was a 242-word passage from ‘Streets for All: Yorkshire’ guidance, which spans 7 pages and contains only 1,529 words including front and back matter \cite{heyorkshire18}.  

\subsection{Data preparation and fine tuning}
After defining specifications to which HAZEL should adhere and identifying a need for fine tuning, we created our datasets and prepared for the process. When this phase of the research began, in late 2023, OpenAI’s GPT models–especially GPT-4–had demonstrated state-of-the-art performance in natural language processing \cite{makyrgiannakis24}. However, at that time, we could only access the beta version of GPT-4, which was available to select stakeholders including the University of Edinburgh. We conducted a brief prompt engineering experiment to compare performance, ultimately deciding to use GPT 3.5-turbo. This decision was based on GPT-3.5-turbo’s greater stability and reduced hallucination tendencies compared to GPT-4.0, which, during beta testing, showed erratic behaviour such as unexpectedly switching languages (see section 5.2). 

We followed the process for fine-tuning delineated in OpenAI documentation \cite{openft24} and resolved issues with support from the platform’s community forum. Here, developers working with OpenAI products share insights, raise questions, and exchange knowledge \cite{opencomm23}. This resource was particularly helpful as we found limited information about fine-tuning OpenAI products elsewhere online, including on commonly consulted resources such as StackOverflow, Github, or Reddit.
 
The first step for fine-tuning was to create a dataset for training and testing. We first converted the HE guidance corpus from PDF images to machine-readable text files using optical character recognition (OCR), completed with the Python package pytesseract (v0.3.13), a wrapper for Google’s Tesseract-OCR engine which extracts text from images \cite{ppitess24}. The resulting text was minimally cleaned to remove whitespace and line breaks, then saved as 237 .txt files in the project directory. 

Though the HE corpus is openly available online, it is currently only accessible to the public as PDF images, meaning that its content is not scrapable or machine readable. Therefore, to mitigate any risks related to the data privacy and security concerns that LLMs have introduced \cite{sebastian23}, \cite{yao24}, \cite{he25} we set up an API key under a zero-data retention (ZDR) agreement with OpenAI through EDINA, a centre for digital expertise and AI infrastructure within the University of Edinburgh’s Information Services. The ZDR agreement, which included a questionnaire about our use case, ensured that no input data would be stored, reused, or accessed by OpenAI during any stage of development or implementation \cite{openent25}.  

We then randomly sampled the .txt corpus without replacement to extract 150 excerpts of approximately 250-300 words each. Each sample in the training data contained only full sentences to allow for estimating readability using formulas, which rely on metrics such as the number of words per sentence. We saved these excerpts in a .csv file and, with assistance from volunteers at HE, manually revised each excerpt to reflect HAZEL’s desired response and correction. The corrected excerpts were saved in a new column in the .csv file.

Short excerpts of longer guidance documents were selected for several reasons. First, creating a dataset with standardised sample lengths allowed us to reliably measure readability using the four formulas. While readability formulas can be calculated on passages of any length, there is a precedent for drawing comparisons between texts of a similar length to ensure accuracy \cite{goltz64}, \cite{fry68}, \cite{lei16}. Second, randomly sampling guidance allowed us to create a training dataset representative of a full piece of guidance, which will often include introductory and concluding sections along with front and back matter. Third, we hoped to minimise costs associated with this research and were financially responsible for tokens input and output during all stages of this process. 

The training and testing data was then converted from a .csv format to JSON Lines as required for fine-tuning with OpenAI’s API \cite{openft24}. Each sample was structured as follows: 
 
{“messages”: [{“role”: “system”, “content”: “message”}, {“role”: “user”, “content”: “message”}, {“role”: “assistant”, “content”: “message”}]} 
 
where the ‘system’ message defines the model’s behaviour, often by assigning it a persona, an identity, or a name; the ‘user’ message represents the input prompt; and the ‘assistant’ message corresponds to the model’s output. In the training data, we constructed the ‘assistant’ message to represent the ideal response to the input prompt for each sample. The training datasets were created to reflect the types of language and changes HAZEL needed to make, which we had agreed upon with the project team and are listed in section 3.1 of this paper. Our system message gave HAZEL its identity as ‘an AI assistant designed to support authors of heritage guidance with writing clear, accessible content for a general audience in the UK’. Each user message included a prompt requesting assistance with a writing task (e.g. revising to the active voice or simplifying language) while the assistant message reflected the ideal output of improved content. 
 
The process for fine tuning required initialising a new OpenAI session with our API key, importing the training data, and creating a new fine-tuning job by selecting a model and setting optional parameters. We completed these steps with v1.57.0 of the Python OpenAI API library \cite{ppiopen24}. Our training dataset contained 80\% of the randomly sampled data structured as messages in JSON Lines format. The parameters we adjusted included the number of training epochs, or the number of passes through the training data during the training session; the temperature, which sets the randomness of the model’s output; and the batch size, which sets the number of samples the model examines at once \cite{ruv23}. 
       
\subsection{Quantitative Evaluation}
Upon completion of fine-tuning, we tested the LLM’s performance. Our testing dataset contained 20\% of the samples that had been reserved from the original dataset, according to the training-testing split recommended in the OpenAI developer forum \cite{opencomm23}. We then examined the results quantitatively, through automated and manual analysis. For automated quantitative testing, we used the Python packages Numpy v2.1.3 \cite{ppinumpy24} and Readability v0.3.2 \cite{ppiread25} to apply four readability formulas to each sample after HAZEL’s revision. These formulas included Flesch-Kincaid, Flesch Reading Ease, Automated Readability Index (ARI), and the Dale-Chall formula. We also calculated the mean, median, and standard deviation of these results aggregated by formula. To determine a baseline for contextualising the scores as a measure of HAZEL’s performance, we calculated readability and the mean, median, and standard deviation for 150 additional excerpts randomly sampled from the HE guidance corpus. The results of these calculations can be found in Table 1 and Table 2. 
  
Each readability formula examines quantifiable features of text to estimate reading difficulty. These metrics include the average sentence length (ASL), the average number of syllables per word (ASW), the average number of characters per word (ACW), the average number of words per sentence (AWS), and the percentage of words that are classified as ‘difficult’ according to an established lexicon (PDW). More specifically, these formulas are calculated as follows:

\begin{description}
    \item[Flesch-Kincaid Grade Level]  
    \[
        FKGL = (0.39 \times ASL) + (11.8 \times ASW) - 15.59
    \]
    where the result corresponds to the reading level required to understand the text in terms of years of American schooling \cite{flesch48}. \\

    \item[Flesch Reading Ease]  
    \[
        FRE = 206.835 - (1.015 \times ASL) - (84.6 \times ASW)
    \]
    where the result is a number from 0–100. The higher the number, the more readable the text; for instance, texts landing in the 50–60 range are considered ‘standard’ in difficulty, while those ranging from 0–30 are classed as ‘scientific’ \cite{flesch48}. \\

    \item[Automated Readability Index (ARI)]  
    \[
        ARI = (4.71 \times C/W) + (0.5 \times W/S) - 21.43
    \]
    where the result corresponds to the reading level required to understand the text in terms of years of American schooling \cite{smith67}.\\  

    \item[Dale–Chall Readability Formula]  
    \[
        Score = (0.1579 \times PDW) + (0.0496 \times ASL) + 3.6365 \quad (\text{if } PDW > 5\%)
    \]
    where PDW stands for the ‘percentage of difficult words’, based on the presence of keywords in a lexicon of 3,000 words deemed easily understandable for 80\% of American fourth-grade students. In cases where the PDW is greater than 5\%, add 3.6365. The higher the total score, the more difficult the text is to read, with a score between 9–9.9 corresponding to a university reading level \cite{dale49}.
\end{description}

In addition to assessing HAZEL in an automated way, we sought the assistance of a copyeditor with expertise in HE’s guidance literature who was therefore able to manually assess the samples qualitatively. The copyeditor had no prior experience with GenAI technologies, meaning that their evaluation would not be influenced by pre-existing ideas. We informed the copyeditor that some of the texts were generated—in whole or in part—by a GenAI system, but did not specify which. All samples included in the copyediting evaluation were written or revised by either HAZEL or ChatGPT to allow for a comparison between the base model and the fine-tuned LLM.

\subsection{Qualitative Evaluation}
The copyeditor was asked to score each sample according to a standardised rubric, which measured five categories on a scale of 1 to 5 to match the criteria agreed with HE staff: readability, concision, clarity, professional tone, formatted and writing for accessibility, and consistent with HE house style guidelines and brand values (see section 3.1.). The samples were divided into five categories, each containing a task related to the guidance writing process for which an author might seek assistance from HAZEL, including copyediting, completing a specific revision, and comparing two texts. These included evaluating text for its overall suitability for HE guidance, comparing two revised versions of text, assessing the quality of an unspecified revision of text, and examining the quality of a targeted revision (e.g. ‘use plain English and avoid jargon’ or ‘use short sentences’). The standard deviation, median, and mean were calculated for the rubric scores, aggregated by chatbot. We also calculated the readability scores for each set of samples. The results of these calculations can be found in Table 3 and Table 4.

Together, this quantitative-qualitative testing protocol was designed to assess how well HAZEL met the accessibility and readability guidelines co-developed with Historic England’s Content team. By comparing HAZEL’s outputs to those of the base ChatGPT model, we aimed to evaluate the extent to which fine-tuning improved the readability and overall suitability of the guidance texts. The following section presents the results of this evaluation. We then discuss their implications for future use of GenAI in heritage communication.
 
\section{Results}
\label{sec:results}
\subsection{‘Interview’ with ChatGPT}
To evaluate whether a general-purpose LLM like ChatGPT could meet Historic England’s writing standards without further adaptation, we conducted a qualitative assessment of its outputs in response to a series of inputs. Consistent with existing research, we found through qualitative review of the ‘transcripts’ of our interactions with ChatGPT (see Supplementary Information) that the chatbot generated false references and misleading content. For instance, when prompted to generate a bibliography on heritage preservation and wildlife, the chatbot provided false titles of articles, journals, and authors. The chatbot also displayed an awareness of works under copyright protection, such as academic monographs, that raise questions about whether its training data includes information under copyright \cite{kandeel24}.  

Furthermore, we identified several areas in which ChatGPT generated text that was inconsistent with the guidelines to which it was instructed to adhere. Surprisingly, in light of existing research suggesting LLM’s suitability for convergent thinking tasks with a discrete ‘right’ answer \cite{castgonz22}, ChatGPT did not perform accurately when prompted to make mechanical changes in text, such as revising to use the active voice or British spelling. Although ChatGPT did correct some well-known spelling and syntactical differences between American and British English (e.g., `color' versus `colour'), it seemed to overlook others, as shown below:

\begin{tcolorbox}[colback=gray!10, colframe=gray!50, left=2mm, right=2mm]
\textbf{Input:} Can you rephrase this in the active voice using British English spelling and grammar? 
The lazy dog was jumped over by the quick brown fox who was carrying a grocery bag full of zucchini and eggplant. 
The fox was yelled at by the orange-colored sun, "Have you remembered to put your sunscreen on? It's nearly noon and it's time for me to burn bright!" \\

\textbf{Output:} The quick brown fox, carrying a grocery bag full of courgettes and aubergines, jumped over the lazy dog. 
The orange-coloured sun yelled at the fox, "Have you remembered to put on your sunscreen? It's nearly noon, and I'm about to shine brightly!"
\end{tcolorbox}

In this exchange, ChatGPT retains the Americanisms ‘grocery bag’, ‘noon’, and ‘sunscreen’ rather than suggesting ‘shopping bag’, ‘midday’, and ‘suncream’, the terms more commonly used in British English. Yet it does recognise the lexical borrowing differences, correcting ‘zucchini’ and ‘eggplant’ to ‘courgettes’ and ‘aubergines’. 

On several occasions, we noticed a failure in prompt chaining leading to what we might call conversational amnesia–in other words, instances where ChatGPT ‘forgot’ the context of a conversation. For example, the following exchange took place after ChatGPT had completed a prompt requisition a revision of text according to a list of criteria:

\begin{tcolorbox}[colback=gray!10, colframe=gray!50, left=2mm, right=2mm]
\textbf{Input:} Gutters and drains need to be checked regularly. \\

\textbf{Output:} I agree! Regular maintenance of gutters and drains is important to prevent blockages and water damage.

\end{tcolorbox}

When working with the base GPT-3.5 turbo model, we observed occasional inconsistencies and errors in its outputs. One notable issue was the model’s limited memory across conversational turns, or breakdowns in chain-of-thought between prompts. In such cases, the model failed to retain or respond appropriately to information provided earlier in the exchange. For example, in one chat session, we instructed the model to revise a series of sentences for improved clarity and conciseness. After completing several of these revisions, we tested its contextual awareness by entering a prompt containing only a sentence: 

\begin{tcolorbox}[colback=gray!10, colframe=gray!50, left=2mm, right=2mm]
\textbf{Input:} Progress remains broadly insufficient to ensure that the buildings sector reaches zero emissions by 2050. \\

\textbf{Output:} Thank you! \\

\textbf{Input:} Sorry, could you revise the sentence? \\

\textbf{Output:} Progress remains insufficient overall to ensure that the buildings sector reaches zero emissions by 2050. 
\end{tcolorbox}

In addition, ChatGPT did not accurately apply common readability formulas, consistently with the findings of previous research \cite{golan23}. As seen in Table 1, ChatGPT underestimated the difficulty of both texts. Although this sample size was small and there are clear scoring differences between Microsoft Word and the readability software, instructing ChatGPT to assess readability once again resulted in variable responses. This suggests that ChatGPT lacks consistent measurement capabilities for these tests. Initially, when asked to describe and then apply the Flesch-Kincaid and Flesch Readability formulas to a given text, ChatGPT provided lower scores. However, when instructed to simply calculate these scores, the results were higher, indicating the text was less readable. Although the limited sample size prohibits drawing any concrete conclusions about the effects of prompt phrasing on the accuracy of LLM-performed calculations, this variability suggests that readability assessments conducted with GenAI or an LLM should be confirmed manually.
 
\begin{table}[h!]
\centering
\caption{Comparison of readability formulas applied to excerpts from Historic England guidance.}
\begin{tabular}{llcccc}
\toprule
Document & Formula & Readable.com & Word & ChatGPT 1 & ChatGPT 2 \\
\midrule
‘3D Laser Scanning’ & Flesch-Kincaid & 13.3 & 14 & 11.82 & 17.07 \\
 & Flesch Readability & 39.4 & 34 & 45.12 & 31.30 \\
‘Streets for All’ & Flesch-Kincaid & 13.1 & 12 & 6.07 & 10.97 \\
 & Flesch Readability & 39.5 & 48.1 & 74.16 & 57.94 \\
\bottomrule
\end{tabular}
\label{tab:readability}
\end{table}

Based on ChatGPT’s inconsistency in executing prompts to correct or generate text and to calculate readability, we determined a need for fine-tuning. 

\subsection{Model selection and fine tuning }
The prompt engineering experiment revealed concerns of so-called ‘hallucinations’—for instance, the model inexplicably switched to German—and we therefore completed fine tuning with the most stable recent release of a GPT model, which was GPT 3.5-turbo. However, we observed that the base model of GPT-3.5 turbo tended to use the passive voice. This remained the case even when the model was explicitly prompted to use the active voice:
\begin{tcolorbox}[colback=gray!10, colframe=gray!50, left=2mm, right=2mm]
\textbf{Input:} How would you revise this sentence? ‘Research on historic department stores was undertaken in 2023.’  \\

\textbf{GPT 3.5-turbo:} In 2023, research was undertaken on historic department stores. \\

\textbf{Input:} Please rephrase this sentence in the active voice: ‘There were more than five thousand applications for listed building consent last year.’ \\

\textbf{GPT 3.5-turbo:} More than five thousand applications for listed building consent were received last year. 
\end{tcolorbox}

Despite following notes on fine tuning available in the OpenAI documentation, we faced challenges during the fine-tuning process and found it necessary to complete several iterations, adjusting the training data and temperature parameter each time, before GPT-3.5 turbo began to behave as HAZEL.  Our initial observations from working with HAZEL, such as how calling HAZEL by name appears to improve the quality and content of the outputs, are summarised in the project’s GitHub page \cite{github24}. We hope that this guide not only informs those working with HAZEL during beta testing, but also serves as a resource for other researchers and developers seeking to adapt OpenAI models to their own aims.

\subsection{Readability scores}
Table 2 summarises the readability scores calculated for three sets of samples: 1) a random sample of unaltered excerpts from the Historic England guidance corpus, 2) guidance edited or generated with ChatGPT, and 3) guidance edited or generated with HAZEL. For each sample, we applied four frequently used readability formulas: Flesch-Kincaid, Flesch Readability, Automated Readability Index (ARI), and Dale-Chall Readability Formula. We also calculated the mean, median, and standard deviation for each readability score, aggregated by sample.

\begin{table}[h!]
\centering
\caption{Mean, median, and standard deviation for four readability formulas calculated for a random sample of excerpts from the Historic England guidance corpus, texts edited with ChatGPT, and texts edited with HAZEL.}
\begin{tabular}{lcccc}
\toprule
Sample & Flesch-Kincaid & Flesch Readability & ARI & Dale-Chall \\
\midrule
\textbf{Historic England corpus} & & & & \\
Mean & 15.47 & 30.63 & 16.98 & 11.50 \\
Median & 15.00 & 31.73 & 16.32 & 11.44 \\
Standard deviation & 3.14 & 12.62 & 3.82 & 1.05 \\
\midrule
\textbf{ChatGPT} & & & & \\
Mean & 13.49 & 31.62 & 14.75 & 10.80 \\
Median & 13.74 & 32.46 & 14.91 & 10.87 \\
Standard deviation & 2.93 & 2.94 & 3.37 & 1.10 \\
\midrule
\textbf{HAZEL} & & & & \\
Mean & 13.20 & 37.15 & 14.15 & 10.65 \\
Median & 12.42 & 38.32 & 12.62 & 10.85 \\
Standard deviation & 0.83 & 0.93 & 0.68 & 0.54 \\
\bottomrule
\end{tabular}
\label{tab:readability_summary}
\end{table}

Both ChatGPT and HAZEL succeeded in slightly lowering the average readability scores across all four formulas. HAZEL’s revisions were scored, on average, as slightly more readable than ChatGPT’s, although still ‘difficult’, meaning more revisions would be required for a general audience. While the magnitude of change across all four formulas is modest—generally less than two grade levels—the reduction in standard deviation for HAZEL’s outputs suggests greater consistency in its revisions. This consistency is particularly valuable for public sector organisations like Historic England, where predictability in tone and clarity across documents supports user trust and accessibility.

\subsection{Copyediting}
The copyeditor evaluated 25 text samples, each originally sourced from Historic England guidance and later revised by either ChatGPT or HAZEL. These samples were assessed against a rubric. The results, summarised in Table 3, show that ChatGPT and HAZEL perform similarly overall, with only slight differences. 

\begin{table}[H]
\centering
\caption{Mean, median, and standard deviation for the rubric scores assigned to each sample by the copyeditor.}
\begin{tabular}{lccccc}
\toprule
Sample & \makecell{Style \\ \& Tone} & \makecell{Clarity} & \makecell{Readability \\ \& Accessibility} & \makecell{Diversity \\ \& Inclusion} & \makecell{Overall \\ Suitability} \\
\midrule
\textbf{HAZEL-produced} & & & & & \\
Mean & 3.8 & 3.73 & 4.07 & 3.57 & 3.57 \\
Median & 4.0 & 4.0 & 4.0 & 4.0 & 4.0 \\
Standard deviation & 0.83 & 0.93 & 0.68 & 0.54 & 0.75 \\
\midrule
\textbf{ChatGPT-produced} & & & & & \\
Mean & 3.81 & 4.0 & 3.89 & 3.87 & 3.62 \\
Median & 4.0 & 4.0 & 4.0 & 4.0 & 4.0 \\
Standard deviation & 0.95 & 0.96 & 0.88 & 1.03 & 0.92 \\
\bottomrule
\end{tabular}
\label{tab:copyediting_scores}
\end{table}

On average, compared to ChatGPT’s revisions, the copyeditor scored the HAZEL-revised samples as slightly more readable and accessible. Although mean scores differed by only small margins (e.g. ±0.2), the lower standard deviation in HAZEL's outputs again indicates more uniform performance. This suggests that, while ChatGPT occasionally produces strong revisions, HAZEL is more reliable in meeting the baseline expectations set by Historic England’s style and accessibility criteria.

\section{Discussion}
\label{sec:discussion}
Overall, our evaluation of ChatGPT and the fine-tuned HAZEL model demonstrates both the promise and limitations of using GenAI for heritage writing tasks. The qualitative prompt engineering experiment with ChatGPT revealed persistent patterns consistent with prior research on LLMs, notably the frequent generation of ‘ghost bibliographic citations’, or references to papers, journals, and monographs that do not exist \cite{orduna23}. These fabricated citations are often presented alongside plausible summaries or abstracts, which makes them appear credible despite being entirely fictitious. Although our prompt focused on bibliographies related to research topics in heritage studies, similar phenomena have been documented in other fields including economics \cite{buchanan24}, medicine \cite{gravel23}, and geography \cite{day23}, indicating that ghost citations are a widespread issue.

This behaviour raises significant concerns for the context of the writing and dissemination of heritage guidance texts. For example, HE guidance documents typically include comprehensive bibliographies to direct readers to additional authoritative sources on a topic. The ‘Adapting Traditional Farm Buildings’ guidance document \cite{hefarm17}, for instance, lists 45 references in its appendix in addition to in-text references throughout the piece. In this context, ghost citations could not only frustrate readers but also undermine the public’s trust in HE guidance and the reputation of the organisation overall. 

Like citation inaccuracies, ChatGPT’s inconsistent adherence to language preferences in our study is in step with similar trends documented before. Prior research attributes LLMs’ English-language bias to the predominance of English texts on the internet \cite{ferrara23}; our observations further suggest a specific bias towards American English embedded within LLM training data. For our purposes, this presented a significant challenge, as maintaining British English standards is essential for accessibility and clarity for HE’s primarily English audience. Similar challenges emerged in the model’s mechanical text edits and readability formula calculations (see Section 4.1). These findings point to inherent limitations in LLM output reliability likely emerging from both the training data and the underlying text generation architecture. 

Fine-tuning addressed many of these issues and resulted in the development of HAZEL, a GenAI chatbot more consistently aligned with Historic England’s content standards. However, this was not without challenges, which we found to align with known difficulties of adapting proprietary LLMs for specific purposes \cite{opencomm23}, \cite{liang23}. One such difficulty consists in the fact that model training data biases can embed certain linguistic patterns \cite{bender21}– e.g., the use of passive voice –making it difficult to correct them by prompt engineering alone. Furthermore, fine-tuning GPT 3.5-turbo to behave as HAZEL required several attempts because, at the time of development, limited documentation on fine-tuning OpenAI models was available beyond community forum \cite{opencomm23}. Throughout this process, we adjusted the training data, system message, and settings such as temperature to guide the model toward outputs consistent with our criteria for accessibility and readability in addition to HE’s tone of voice and style. 

During testing, we observed that prompt engineering was still an essential step in working with HAZEL. For example, we found that explicitly referring to the model by name—e.g. ”HAZEL, can you help me with a piece of guidance I am working on?”—often improved the clarity, relevance, and quality of its outputs. As such, we conceptualise prompt engineering and fine-tuning as a symbiotic process, with the results of prompt engineering informing our approach to fine-tuning (and vice-versa). Though more research would be needed to generalise this observation to other contexts and to other LLMs, researchers working to fine-tune GPT models for writing tasks might find that their results benefit from a similar iterative approach. This is perhaps especially the case when the desired outputs of the model conflict with its ‘default’ behaviour, such as when generated text is desired in an English-language dialect other than Standard American English.

Our quantitative results (Table 2) demonstrate that fine-tuning LLMs can improve how closely generated text meets desired criteria, leading to modest improvements in readability for a general audience based on four commonly used readability formulas. According to these metrics, HE guidance materials revised by HAZEL were, on average, slightly more readable than those revised by ChatGPT. HAZEL-revised texts also showed a lower standard deviation in readability scores compared to ChatGPT revisions, suggesting that HAZEL produced more consistent readability. Revising written text into plain language is a commonly proposed strategy to enhance accessibility \cite{boldyreff01}, \cite{sedgwick21}. However, some guidance documents published by HE are highly technical, making it challenging to simplify them without losing meaning. Although both HAZEL and ChatGPT improved readability scores compared to the original corpus, most texts remained classified as ‘difficult’ to ‘very difficult’ to read. One limitation of our study is that the HE guidance corpus was sampled randomly, which may have resulted in an overrepresentation of highly specialized, technical guidance pieces that are inherently challenging to simplify without missing critical detail. 

Potential solutions include reshaping glossary sections in PDF documents into clickable embedded definitions, enabling readers to access explanations without interrupting their reading flow. Another approach is providing plain language summaries that offer a concise overview, with expandable sections for deeper exploration—a feature already partially implemented on the HE website for some heritage guidance.  However, interactive web content introduces new challenges \cite{rubano20}, \cite{bhatia24}, such as accessibility issues for users of assistive technologies or those reading on mobile or tablet devices.

To address these complexities, future development could explore alternative LLMs or approaches such as retrieval-augmented generation, which integrates external knowledge sources to enhance both readability and factual accuracy. A beta testing phase integrating qualitative user feedback into the design process would better ensure that HAZEL–and similar technologies–effectively support authors of heritage guidance throughout the writing process.

Finally, our study findings showed a trade-off between readability and clarity. Although, on average, ChatGPT performed less well on readability, it produced slightly clearer revisions that were more aligned with HE style than HAZEL. In particular, both models scored lowest on ‘Overall Suitability’, with HAZEL’s ‘Diversity and Inclusion’ score matching this low rating while outperforming ChatGPT on ‘Readability \& Accessibility’, the primary target of our fine-tuning. This indicates that neither system consistently meets HE’s expectations for guidance literature. In other words, ChatGPT and HAZEL’s ability to generate what we would call ‘good writing’ is not equivalent to high-quality work. The repetitive, similar flaws observed also suggest a lack of variety in the style or word choice of the revised texts. Though the size of this study is too small to support broader conclusions, we wish to highlight that the copyeditor’s observations are consistent with other observations about ChatGPT’s similarity in writing style in creative writing tasks \cite{doshi23}. If future work is able to confirm that ChatGPT indeed exhibits a particular writing style, at least in between updates, then care needs to be taken where GenAI is deployed in heritage settings in order to retain the diversity of voices and perspectives in published materials. 
\subsection{Limitations}
 Our study was subject to several limitations. First, copyediting feedback was sourced from a single expert. This is because it was essential for the copyeditors to be experts in HE guidance with no prior experience using GenAI tools, significantly limiting the number of professionals we could recruit. Second, time and funding constraints limited the study’s scope. Because HAZEL was built on GPT 3.-5-turbo—an OpenAI product requiring a paid subscription—significant testing or public release of HAZEL was not feasible. As a result, we were unable to conduct formal beta testing or observe its deployment in live professional settings. Evaluating HAZEL’s effectiveness in real-world workflows, such as outlining guidance documents or producing summaries, remains a key area for future research.
 
Additionally, we did not directly examine potential affective dimensions of (Gen)AI, which are well established as a critical area of study in the field of human–computer interaction \cite{wilson10}, \cite{montag25}. Dr Susan Hazan, a professional in the museum sector, articulates a profound unease about GenAI, recounting her experience when asking ChatGPT to emulate her writing style:

\begin{tcolorbox}[colback=gray!10, colframe=gray!50, left=2mm, right=2mm]
“I felt that something had got into my mind and was scraping my brain…this ‘thing’ was crawling around my published chapters and papers and other corners of the Internet and presuming to come out with the conclusions I would come to myself based on what it learned about me. My ideas had become a collage of presumptions that, when stringed together spewed out a convincing doppelganger of me.” \cite{hazan20}
\end{tcolorbox}

This response is reminiscent of the uncanny valley \cite{mori12}, or the sense of unease when encountering machines that seem almost-but-not-quite human, and highlights negative, subjective experiences of GenAI that remain underexplored in heritage settings. Such fears may stem from a lack of familiarity with AI tools or prior negative encounters with digital technologies, but they coexist with tangible ethical concerns including bias and authorship. In the context of this study, we acknowledge that negative affect toward GenAI could act as a barrier to its adoption in the heritage sector, though additional research is needed to explore this further.

The study did not evaluate the integration of GenAI tools into long-term collaborative heritage workflows. Future research should explore HAZEL’s impact on productivity, trust, and cognitive load when embedded into existing authoring processes. GenAI intersects in important ways with FAIR (Findable, Accessible, Interoperable, and Reusable) data principles and the broader values of open science. While AI-generated outputs may enhance documentation and dissemination, their training data and structure remains a ‘black box’. As recent lawsuits have shown \cite{veltman25}, \cite{allyn24}, at least some models are trained on massive web-scraped datasets without clear licensing or copyright clearance. There is also a risk that GenAI could perpetuate existing cultural and linguistic biases embedded in the training data, undermining efforts to democratise heritage texts.

Additionally, the use of proprietary or subscription-only models, such as GPT models, presents barriers to equitable participation. Significant ethical concerns aside, many cultural heritage organisations—particularly smaller and under-resourced institutions—lack the financial resources to sustain paid access to commercial AI platforms. The ethical use of GenAI in heritage must therefore prioritise open-access, freely available models. Professionals who would be working with GenAI would also benefit from onboarding and training modules to help them use these tools responsibility while mitigating potential risks of, for example, information fabrication. 

\section{Conclusions}
This study provides a timely and critical assessment of generative AI’s (GenAI) potential in the context of heritage writing focusing on applications within institutions such as Historic England (HE). Our findings reveal both encouraging opportunities and significant risks involved in integrating GenAI tools—both general-purpose chatbots such as ChatGPT and fine-tuned chatbots such as HAZEL—into heritage writing processes. GenAI shows clear potential to augment writing tasks in the sector, particularly in improving the readability of complex technical texts and assisting with surface-level copyediting. When fine-tuned, these tools can simplify language and support consistency of tone and structure, improving readability and accessibility of heritage guidance documents. These capabilities hold promise for expanding participation in heritage conservation and aligning outputs with the goals of cultural democracy.

However, these benefits must be weighed against notable limitations. GenAI systems, regardless of training or tuning, lack lived experience, contextual awareness, professional expertise, and cultural sensitivity required for creating many heritage guidance texts. Therefore, these systems might be best limited to convergent thinking tasks in the writing process, such as proofreading and other tasks with discrete options. In divergent tasks requiring creative problem-solving and subjective expertise, GenAI writing assistants introduce risks such as stylistic flattening, fabrication of information, generic phrasing, and at times, brand or value misalignment. Building on the study findings, several directions for future work emerge. First, HAZEL should be evaluated in the context of the heritage writing process, such as collaborative drafting or iterative editing of real guidance texts. Observing user experiences, barriers, and outcomes will offer a fuller picture of HAZEL’s performance and suitability. Second, the emotional and cognitive dimensions of GenAI use in heritage—such as authorship anxiety, trust, and perceptions of automation—could be examined in more depth. These affective factors may be as consequential as technical accuracy in shaping adoption and responsible use. 

Third, fine-tuning methods should be expanded and other methods explored. Retrieval-augmented generation (RAG) and other hybrid techniques may improve adherence to HE style and tone of voice guidelines while reducing the risk of errors or value misalignment. Co-designing these models with HE professionals and accessibility specialists would further enhance their effectiveness and ethical fit. Finally, broader ethical, legal, and policy frameworks must be developed to guide GenAI integration in the heritage sector.

In conclusion, fine-tuned GenAI tools like HAZEL offer promise in supporting efforts to make heritage texts more readable and accessible. However, due to significant potential risks, any integration of GenAI in the heritage sector will require caution. 

\section{Acknowledgments}
This study was funded by Historic England’s Digital Strategy Board Innovation Fund. The funder had no role in the design of the study, data collection, analysis or interpretation of the data, or in the writing of this manuscript. The authors would like to thank James Reid (EDINA, University of Edinburgh) for establishing the zero data retention agreement with OpenAI and for supporting access to GPT models through the University’s digital research infrastructure. We also thank David Jones for his copyediting expertise during the qualitative portion of the study. Additionally, we gratefully acknowledge Daniel Pett (Historic England) for facilitating the collaboration of the project team and assisting with project design and funding application.

\bibliographystyle{plainnat}

\end{document}